\documentstyle[prb,aps,epsfig]{revtex}
\begin{document}
\newcommand{\be}{\begin{equation}}
\newcommand{\ee}{\end{equation}}
\newcommand{\ba}{\begin{eqnarray}}
\newcommand{\ea}{\end{eqnarray}}
\newcommand{\Ln}{\Lambda_n}
\newcommand{\Lc}{\Lambda_c}
\newcommand{\dpdx}{ {{d\phi} \over dx}}
\newcommand{\dpdt}{ {{d\phi} \over dt}}
\newcommand{\dedx}{ {{d\eta} \over dx}}
\newcommand{\dedt}{ {{d\eta} \over dt}}
\newcommand{\adag}{a^{\dag}}
\newcommand{\bdag}{b^\dag}
\title{Edge Occupation Numbers of a Half-Filled Landau Level}
\author{J. H. Han$^{1}$ and S.-R. Eric Yang$^{1,2}$}
\address{Asia Pacific Center for Theoretical Physics, 207-43
   Cheongryangri-dong Dongdaemun-gu Seoul 130-012, Korea$^{1}$
        \\ and \\
  Department of Physics, Korea University, Seoul 136-701,
    Korea$^{2}$}

\maketitle
\draft

\begin{abstract}
We demonstrate that a theory of the edge of a half-filled Landau level
recently proposed by Lee and Wen predicts results for the edge occupation
number similar to those of a variational trial wave function proposed
previously by us.  We treat Lee and Wen's edge action of a half-filled
Landau level within the framework of  bosonization theory, and show
that the momentum occupation numbers are determined by a product of two
Green's functions, one charged and one neutral. In the bulk region ($k<0$)
we find a linear occupation profile, $n_k \propto A+Bk$, while in the tail
region ($k>0$) it is exponentially decaying over the range $k\sim\Ln$,
the momentum cutoff for neutral mode. We find a good fit with the
numerical results for occupation numbers.
\end{abstract}

Recently there has been a great deal of interests in two-dimensional
electron liquids in a strong magnetic field at half-filling, and
several theories have been proposed
\cite{halp,read1,shankar,hal,yang1,lee,read2}.
Unlike insulating QH liquids this new state is metallic.  Although
all these theories describe a metallic state it is
not clear whether they predict identical edge properties.
Field theoretical methods of Shytov {\it et al.} \cite{shy} and Lee and
Wen\cite{leewen} seem to predict
different dynamic properties of the edge.
It is also not clear whether the edge properties of a variational
trial wave function of
this liquid \cite{yang1} differs from those predicted by field theories.

After attaching two flux quanta to each electron to realize the
Chern-Simons transformation at half-filling, one is apparently left
with a Fermi sea without external magnetic field. One can alternatively
regard it as $N/2$ of the composite fermions seeing an upward $N/2$
quanta of flux, while the other half sees magnetic flux in the opposite
direction, in an $N$-particle system. On average the net residual flux
will be zero.  As a consequence half of the composite fermions in the
system occupy the ``lowest Landau level" (as judged by the composite
fermions), and the other half the anti-holomorphic lowest Landau level.
This consideration naturally leads to a trial wave function at
half-filling in particular, and wave functions at other hierarchy
filling factors in general\cite{yang2,yang1}.

Once a trial wave function for the half-filled Landau level
\cite{yang1} is known the edge occupation number can be computed in the
same manner as the calculation of the occupation number from Laughlin's
wave function\cite{occ}.  It turned out that the edge of a
metallic liquid has a rather different property than that of
incompressible QH states at filling factor $\nu=1/M$ ($M$ odd
integer).  It is well known that for incompressible QH states at
$\nu=1/M$ the electron occupation number $n_k$ near the edge is $n_{k}
\propto (k_{ed}-k)^{M-1}\theta(k_{ed}-k)$, where $k_{ed}$ is the
wavenumber corresponding to the location of the edge\cite{wen,occ}. 
Our result for the quantum Hall droplet at
half filling showed that a tail exists in the occupation number beyond
the radius of a droplet of uniform filling factor $1/2$, i.e. $n_{k}
\not=0$ for $k-k_{ed}>0$.  This apparent increase in the size of the
droplet is due to the displaced zeroes of wave function from the
electrons \cite{yang2}.  The occupation numbers were found to exhibit
Friedel type oscillations, consistent with the metallic nature of the
liquid\cite{yang2}.

Recently Lee and Wen have proposed a two-mode model for the edge \cite
{leewen} based on a theory of the electron liquid at $\nu=1/2$
consisting of charge and flux liquids \cite{lee}.  In this theory the
charge and the neutral modes of edge deformations are assumed to be
well defined up to the momentum cutoffs $\Lambda_{c,n}$, beyond which
modes become over-damped.  Lee and Wen have calculated tunneling
coefficient using this theory, and have found that it agrees with the
experimental data of Grayson {\it et al.}\cite{gray}

In this paper we demonstrate that Lee and Wen's theory predicts the
edge occupation number profile similar to what is found from the trial
wave function\cite{yang1}.  We treat their proposed action within the
framework of bosonization theory and find that $n_{k}\propto A+Bk$ for
$-\Lambda_{c}\ll k<0$ and $n_{k}\propto e^{-k/\Ln}$ for $k>0$ ($k$ is
measured from $k_{ed}$).  Our trial wave function also gives linearly
increasing occupation numbers for small negative $k$.

When a proper action is given for the edge at half-filing the electron
Green's function can be calclulated.
The edge action of a half-filled Landau level derived from the 
dipole theory of Lee\cite{lee}  is
\be
S_e = {1\over 2\pi}\int \dpdx \left(\dpdt\!-\!v_c \dpdx \right)dxdt-
      {1\over 4\pi}\int \dedx \left(\dedt+v_n \dedx\right)dxdt+
      \int (2\phi\!-\!\eta)(\vec{j}_t . \hat{n} -\partial_\mu j^{\mu}) dxdt.
\label{eq:edge_action}
\ee
It is important to note that the neutral mode is counter-propagating
compared with the charged mode, as a result of the minus sign that comes
in the $\partial_t \eta\partial_x \eta$ part of the action and 
the positive-definiteness of
the Hamiltonian. The last term in Eq.\ (\ref{eq:edge_action}) is the
coupling of the edge
mode with the transverse bulk current ($\vec{j}_t .\hat{n}$) or the tunneling
($\partial_\mu j^{\mu}$) current. We parameterize the sample boundary with $x$,
using $\hat{n}$ to indicate the outward normal to the  boundary.
We quantize the edge
action in the absence of the coupling to current sources. The canonical
commutation
relations for the fields $\phi$ and $\eta$ are
\be
[\dpdx, \phi (y)]=\pi i \delta (x-y); \,\,\,\,
[\dedx, \eta (y)]=-2\pi i \delta (x-y).
\ee
By comparison with the chiral Luttinger liquid theory at other filling
factors\cite{wen,han}, one may associate $2d\phi /dx$ with the charge
displacement at
the edge of the liquid. Since the one-dimensional density is
$(\nu/2\pi)$(displacement), electron creation/annihilation operators at the boundary
will need to satisfy the following algebra:
\be
[{\nu\over 2\pi}(\mbox{displacement}), (\mbox{electron op.})]=\pm
\delta(x-y)(\mbox{electron op.})
\label{eq:charge_comm}
\ee
One can easily show that $e^{\pm 2i\phi}$ satisfies the desired
algebra, but when one interchanges two such operators at $x$ and $x'$, one
finds that they obey Bose, rather than Fermi, statistics.
To fix the statistics without altering the
relation in Eq.\ (\ref{eq:charge_comm}), we use the following
annihilation/creation operators
\be
\psi(x)\sim e^{-2i\phi(x)\pm i\eta(x)}, \,\,\,
\bar{\psi}(x)\sim e^{2i\phi(x)\mp i\eta(x)}.
\label{eq:bosonized_el}
\ee
In Lee and Wen's paper, a specific combination $2\phi-\eta$
appears in the electron Green's function, while our formulation seem to
favor both signs.
One presumably needs the coupling term in Eq.\ (\ref{eq:edge_action}) in
order to discriminate
the two signs. It will be clear that all the computations below will be
insensitive to the choice of  sign. We
introduce the mode expansion of the boson fields:
\ba
\phi(x)&=&\phi_0 \!+\!f_0 x\!+\!\sum_{n>0}{i\over \sqrt{2n}}
[\adag_n  e^{-inx}\!-\!a_n e^{inx}]; \,\, [f_0, \phi_0]=i/2, \nonumber \\
\eta(x)&=&\eta_0 \!+\!g_0 x\!+\!\sum_{n>0}{i\over \sqrt{n}}
[b_{n}  e^{-inx}\!-\!\bdag_{n} e^{inx} ]; \hspace{0.4cm} [g_0, \eta_0]=-i.
\ea
The ground state is defined by the action of the annihilation operators,
$a_n |gs\rangle=b_n |gs\rangle=0$. The time dependence is given by setting
$x\rightarrow x-v_c t$ for the charged mode, and $x\rightarrow x+v_n t$
for the neutral mode.

Since this is a bosonization theory, the phonon Green's functions
determine the underlying electron Green's function as well. Define
$z\equiv x-v_c t$, and $w \equiv x+v_n t$. We have
\ba
G_\phi (z)&=&4\langle \phi(z) \phi(0)\rangle-
4\langle \phi(0) \phi(0)\rangle=
2\int_0^\infty {dk\over k} [e^{ikz}-1]e^{-k/\Lc}=-2\ln [1-i \Lc z],
\nonumber \\
G_\eta (w)&=&\langle \eta(w) \eta(0)\rangle-
\langle \eta(0) \eta(0)\rangle=-\ln [1+i \Ln w].
\ea
Electron Green's functions are obtained as
\ba
\langle \bar{\psi}(z)\psi(0)\rangle &=&
\langle e^{2i\phi(z)}e^{-2i\phi(0)}\rangle
\langle e^{-i\eta(z)}e^{i\eta(0)}\rangle=\exp[G_\phi (z)+G_\eta (w)],
\nonumber \\
\langle \psi(0)\bar{\psi}(z)\rangle &=&
\langle e^{-2i\phi(0)}e^{2i\phi(z)}\rangle
\langle e^{i\eta(0)}e^{-i\eta(z)}\rangle=
\exp[G_\phi (-z)+G_\eta(-w)].
\ea
It also follows from the definition in Eq.\ (\ref{eq:bosonized_el}) that
$\langle \psi(z)\bar{\psi}(0)\rangle=
\langle \bar{\psi}(z)\psi(0)\rangle$. Putting our results together,
\ba
\langle \psi(z)\bar{\psi}(0)\rangle=
\langle \bar{\psi}(z)\psi(0)\rangle=1/[1-i\Lc z]^2 [1+i \Ln w], \nonumber
\\
\langle \psi(0)\bar{\psi}(z)\rangle=
\langle \bar{\psi}(0)\psi(z)\rangle=1/[1+i\Lc z]^2 [1-i \Ln w].
\ea
The time-ordered Green's function $G_e (x,t)=\theta(t)\langle
\bar{\psi}(z)\psi(0)\rangle-\theta(-t)\langle
\psi(0)\bar{\psi}(z)\rangle$ is given by
\be
G_e (x,t)={\theta(t) \over {[1-i\Lc z]^2 [1+i \Ln w]}}-
         {\theta(-t)\over {[1+i\Lc z]^2 [1-i \Ln w]}}.
\ee
One can consider another kind of
Green's function, $G_b (x,t)=\exp [\theta(t)\{G_\phi (z)+G_\eta
(w)\}+\theta(-t)\{G_\phi (-z)+G_\eta (-w)\}]$:
\be
G_b (x,t)={\theta(t) \over {[1-i\Lc z]^2 [1+i \Ln w]}}+
         {\theta(-t)\over {[1+i\Lc z]^2 [1-i \Ln w]}}.
\ee
This function is clearly bosonic
in character, $G_b (-x,-t)=+G_b (x,t)$. On the other hand, the electron
Green's function
satisfies $G_e (-x,-t)=-G_e (x,t)$. Notice that one has the
approximate symmetry $G_e (x,-t)\approx -G_e (x,t)$ provided $\Lc x, \Ln
v_n t \ll 1$, because we have, in such a limit,
\be
G_e (x,t)\approx {sgn (t)\over t^2}.
\ee
We will assume that the momentum cutoffs satisfy $\Ln \ll \Lc$. For
the bosonic Green's function $G_b (x,t)$, one observes a similar symmetry in the
slightly different limit $\Ln v_n t \ll 1 \ll \Ln x$, and $x\ll v_c t$;
\be
G_b (x,t)\approx {1\over t^2}\left(
{\theta(t) \over {i\Ln x}} -{\theta(-t) \over {i \Ln x}}\right)=
{{sgn (t)/t^2} \over {i \Ln x}}\approx -G_b (x,-t).
\ee
The above expression embodies the symmetry that was extensively
discussed and
emphasized by Lee and Wen\cite{leewen}. It says that the desired statistics
property shows up at distances short
compared with $v_c t$, but larger than $1/\Ln$.

We now consider the static property of the edge, in particular its
momentum occupation. The momentum
distribution is simply a Fourier transform of the equal-time Green's
function $G_e (x,0^+)=G_b (x,0^+)$:
\be
\label{occnm}
n_k \propto \int_{-\infty}^{\infty} {{e^{ikx}\,\,\, dx}\over
{[1-i\Lc x]^2 [1+i \Ln x]}}\propto
-ke^{k/\Lc}\theta(-k)+ {\Ln \Lc \over{\Ln+\Lc}}e^{-k/\Ln}\theta(k).
\ee
Due to the presence of counter-propagating neutral mode, we have an
extra occupation on the $k>0$ side that decays exponentially over the
distance $k\sim \Ln$. We believe the discontinuity at $k=0$ is not
physical and is an artifact of taking the integration over infinite range.
We can instead work with the Green's function for finite circumference 
$L$:

\ba
G_\phi (z)&=&
-2\ln \left(
{{1\!-\!\exp[-2\pi(1\!-\!iz\Lc)/\Lc L]}\over
{1\!-\!\exp[-2\pi/\Lc L]}}\right),
\nonumber \\
G_\eta (w)&=&-\ln \left(
{{1\!-\!\exp[-2\pi(1\!+\!iw\Ln)/\Ln L]}\over
{1\!-\!\exp[-2\pi/\Ln L]}}\right).
\ea
The corresponding occupation number $n_m$ is given by
\be
n_m \propto \int_0^1
{ \exp[2\pi i m x]\,\,\, dx \over
{[1\!-\!\exp(-2\pi (1\!-\!i\kappa_c x)/\kappa_c) ]^2
 [1\!-\!\exp(-2\pi (1\!+\!i\kappa_n x)/\kappa_n) ]}}.
\ee
We have introduced $\kappa_{c,n} =\Lambda_{c,n} L$ and measured $x$ in 
units of $L$.
Expanding the denominators in powers series, one obtains a simple
expression for $n_m$:

\be
n_m \propto \sum_{s=0}^\infty \sum_{t=0}^\infty
\delta_{t-s,m} (s+1)e^{-2\pi s/\kappa_c -2 \pi t/\kappa_n}.
\ee
Normalizing by the value at $m=0$, 
\be
n_m /n_0 =\left\{ \begin{array}{ll}
    e^{-2\pi m/\kappa_n}, & m>0 \\
  e^{2\pi m/\kappa_c}
\{1\!-\!m(1\!-\!e^{-2\pi(\kappa_c^{-1}\!+\!\kappa_n^{-1})})\}, & m<0.
               \end{array} \right.
\label{eq:finite_occnm}
\ee
It is exponentially decaying in the $m>0$ region (tail) and linear, of the
form $A+Bm$, for $-\kappa_c\! \ll\! m<0$. The coefficients A and B
depend on the exact
value of the cutoff parameters. To connect with the thermodynamic limit, one can rewrite the
occupation numbers in terms of $k$. We assume that the momentum cut-offs $\Ln$ and $\Lc$
remain finite in the $L\!\rightarrow\!\infty$ limit.
\be
n_k /n_0 =\left\{ \begin{array}{ll}
    e^{-k/\Ln}, & k>0 \\
  e^{k/\Lc}
(1\!-\!k(\Lc^{-1}\!+\!\Ln^{-1})), & k<0.
               \end{array} \right. 
\label{eq:infinite_occnm}
\ee

Figure 1 shows occupation numbers for $N=11$ electrons at half-filling
obtained from finite-size exact diagonalization\cite{yang1}.  Our trial
wave function indicates that the tail region, $x>0$ with $x=kR_{ed}$,
does not disappear in the thermodynamic limit. Here $R_{ed}=2\sqrt{N}$
is the radius of a uniform, compact half-filled droplet.  The trial
wave function was constructed by occupying about half of the composite
fermions in the lowest Landau level, while each of the other half
singly occupies the higher Landau levels after the projection. In the
Laughlin state all of the composite fermions occupy the lowest Landau
level to form an incompressible liquid\cite{jain}.  Such differences in
the occupation behavior of composite fermions underlie the observed
difference of electron occupation for compressible/incompressible
states.  From Eqs.\ (\ref{eq:finite_occnm})
and (\ref{eq:infinite_occnm}) we
expect $n_k \!=\!A+Bk$ for small negative $k$, and we find that this
expression (open squares) fits the numerical results well for suitable
choice of $A$ and $B$.

In this paper we have demonstrated that the edge occupation numbers
calculated from our trial wave function are similar to those of Lee and
Wen's theory.  It would be interesting to see if 
similar excitation properties can be found.

The work of SREY  has been supported by
the Ministry of Education
under Grant No. BSRI-96-2444.

\begin{figure}[ht]
\centering
\epsfig{file=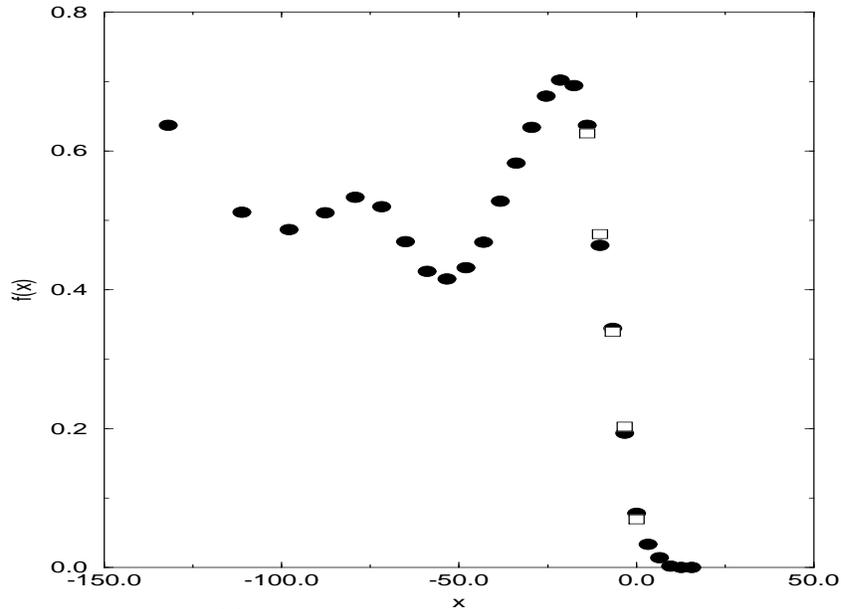,height=8cm,width=11cm}
\caption{Occupation numbers for $N=11 (\bullet)$ computed numerically.
Occupation numbers for small negative values of $x$
are fitted with a linear function (open squares).}
\end{figure}

\end{document}